\documentclass[preprint, nopreprintline, authoryear, lefttitle, 5p]{elsarticle}
\graphicspath{{Figures/}}

\usepackage[english]{babel}

\usepackage{amssymb,amsmath,booktabs,color,dcolumn,bm,thumbpdf}
\DeclareMathOperator{\tr}{\mathrm{tr}} 
\DeclareMathOperator{\expect}{\mathrm{E}}

\usepackage{hyperref}
\definecolor{dred}{rgb}{0.5,0,0}
\definecolor{dblue}{rgb}{0,0,0.5}
\AtBeginDocument{\hypersetup{%
  hyperindex = {true},
  colorlinks = {true},
  linktocpage = {true},
  plainpages = {false},
  linkcolor = {dblue},
  citecolor = {dblue},
  urlcolor = {dred},
  pdfstartview = {Fit},
  pdfpagemode = {UseOutlines},
  pdfview = {XYZ null null null}
}}

\begin{document}

\begin{frontmatter}

\title{Bias reduction as a remedy to the consequences of infinite estimates in Poisson and Tobit regression}

\author[1]{Susanne K\"oll}
\ead{Susanne.Berger@student.uibk.ac.at}
\author[2]{Ioannis Kosmidis}
\ead{Ioannis.Kosmidis@warwick.ac.uk}
\author[3]{Christian Kleiber}
\ead{Christian.Kleiber@unibas.ch}
\author[1]{Achim Zeileis\corref{cor1}}
\ead{Achim.Zeileis@uibk.ac.at}
\address[1]{Faculty of Economics and Statistics, Universit\"at Innsbruck, Austria}
\address[2]{Department of Statistics, University of Warwick \& The Alan Turing Institute, United Kingdom}
\address[3]{Faculty of Business and Economics, Universit\"at Basel, Switzerland}
\cortext[cor1]{Corresponding author}

\begin{abstract}
  Data separation is a well-studied phenomenon that can cause problems
  in the estimation and inference
  from binary response models.
  Complete or quasi-complete separation occurs when
  there is a combination of regressors in the model whose value can
  perfectly predict one or both outcomes. In such cases, and such
  cases only, the maximum likelihood estimates and the corresponding 
  standard errors are infinite. It is less widely known
  that the same can happen in further microeconometric models. 
  One of the few
  works in the area is \cite{bias:SantosSilva+Tenreyro:2010} who note
  that the finiteness of the maximum likelihood estimates in Poisson regression depends on the data configuration
  %that there is positive probability for infinite estimates in Poisson
  %regression
  and propose a strategy to detect and overcome the
  consequences of data separation. However, their approach can lead to
  notable bias on the parameter estimates when the regressors are
  correlated. We illustrate how bias-reducing adjustments to the
  maximum likelihood score equations can overcome the consequences of
  separation in Poisson and Tobit regression models.
\end{abstract}


\end{frontmatter}

\section{Sources of separation in regression models}
\label{sec:intro}
Suppose that the non-negative random variable $y_i$ has a distribution with a point 
mass at zero.\footnote{Note that the discussion here extends to the
  case where the support of the response is bounded below or above. If
  the lower boundary is a constant $\underline{b} \ne 0$, we can use
  $y_i - \underline{b}$. Similarly, if the upper boundary is
  $\overline{b}$, we can use $\overline{b} - y_i$.}  Suppose that the
distribution function of $y_i$ is $F(\cdot; \mu_i, \phi)$
($i = 1, \dots, n$), where the scalar parameter $\mu_i$ is a
centrality measure (e.g., the mean), and the parameter $\phi$
represents higher-order characteristics of the distribution (e.g.,
dispersion).

A regression model can be formulated as
\begin{eqnarray}
\label{dgp1} y_i & \sim & F(\cdot; \mu_i, \phi) \, , \\
\label{dgp2} \mu_i & = & h(x_i^\top \beta) \quad (i = 1,\ldots, n)\, ,
\end{eqnarray}
where $x_i$ is a vector of regressors with
$\mathop{\rm dim}(x_i) = p$, which is observed along with $y_i$, and
$h(\cdot)$ is a monotonically increasing function that links $\mu_i$
to $x_i$ and a parameter vector $\beta$. The model specification
in~(\ref{dgp1}) and~(\ref{dgp2}) covers a range of models, including
models for binary, multinomial, ordinal, and count models,
%stats% \citep[see, e.g.,][]{bias:Winkelmann+Boes:2009}, 
models for limited dependent
variables such as the Tobit model
%stats% \citep[see][]{bias:Tobin:1958}
and its extensions, and zero-inflated and two-part or hurdle models.
%stats% \citep[see][for a review]{bias:Min+Agresti:2002}.

The existence of a point mass at zero implies that
$f(0; \mu_i, \phi) = F(0; \mu_i, \phi)$, where
$f(\cdot; \mu_i, \phi)$ is the density or probability mass function
corresponding to $F(\cdot; \mu_i, \phi)$.

The simplest but arguably often-encountered occurrence of data
separation in practice is when there is a regressor
$x_{i,k} \in \{0, 1\}$ such that $y_i = 0$ for all
$i \in \{1, \ldots, n\}$ with $x_{i,k} = 1$. Assuming that
$y_1, \ldots, y_n$ are independent conditionally on
$x_1, \ldots, x_n$, the log-likelihood $\ell(\beta, \phi)$ for the
model defined by~(\ref{dgp1}) and~(\ref{dgp2}) can be decomposed as
\begin{eqnarray}
% \ell(\beta, \phi) & = & \sum_{i = 1}^n \log f(y_i; h(x_i^\top \beta), \phi) \\ 
\label{loglik1}  \ell(\beta, \phi)                
& = & \sum_{x_{i,k} = 0} \log f(y_i; h(x_{i,-k}^\top \beta_{-k}), \phi) + \\ 
\label{loglik2}                   
& & \sum_{x_{i,k} = 1} \log F(0;   h(x_{i,-k}^\top \beta_{-k} + \beta_{k}), \phi) ,
\end{eqnarray}
where $a_{-k}$ indicates the sub-vector formed from a vector $a$ after
omitting its $k$-th component.

Term~(\ref{loglik1}) is exactly the log-likelihood without the $k$-th
regressor and based only on the observations with $x_{i,k} = 0$. Under
the extra assumption that $F(0; \mu_i, \phi)$ is monotonically
decreasing with $\mu_i$ (which is true, for example, in Poisson
%stats% log-linear
and Tobit regression models), $\beta_k$ will diverge to
$-\infty$ during maximization, so that~(\ref{loglik2}) achieves its
maximum value of $0$. Then, the maximization of term~(\ref{loglik1})
with respect to $\beta_{-k}$ yields the maximum likelihood (ML)
estimate of $\hat\beta_{-k}$. So, the ML estimate of $\beta_{-k}$ will
be the same as the ML estimate obtained by maximizing the
log-likelihood without the $k$-th regressor over the subset of
observations with $x_{i,k} = 0$.

As \cite{bias:SantosSilva+Tenreyro:2010} show for Poisson
regression,
%stats% log-linear models,
the same situation can occur more generally, when
separation occurs for a certain linear combination of regressors. Our
discussion above extends their considerations
beyond log-link models and Poisson regression.
%stats% to models with possibly different distributions than Poisson and more general link functions.

\section{Estimating regression models with separated data}

\cite{bias:Albert+Anderson:1984} showed that infinite estimates in
multinomial logistic regression occur if and only if there is data
separation. Since then, the consequences of infinite estimates to
estimation and inference have been well-studied for binomial and
multinomial responses.

A popular remedy in the statistics literature is to replace the ML
estimator with shrinkage estimators that are guaranteed to take finite
values \citep[see, for example][for using shrinkage priors in the
estimation of binary regression models]{bias:Gelman+etal:2008}. The
probably most-used estimator of this kind comes from the solution of
the bias-reducing adjusted score equations in \citet{bias:Firth:1993}
(see, for example, \citealt{bias:Heinze+Schemper:2002} and
\citealt{bias:Zorn:2005} for accessible detailed accounts),
% \citep[see, for example][for accessible, detailed
% accounts]{bias:Heinze+Schemper:2002, bias:Zorn:2005},
which guarantee
estimators with smaller asymptotic bias than what the ML estimator
typically has \citep{bias:Firth:1993,bias:Kosmidis+Firth:2009}.

In contrast, the majority of methods that have been put forward in the
econometrics literature are typically based on omitting the regressors
that are responsible for the infinite estimates. Such practice can be
problematic as we discuss in the following sections.
% We are also not aware of viable proposals for limited dependent
% variable models, such as Tobit regression.

% Below, we illustrate the potential problems that can occur by merely
% omitting covariates in Poisson and Tobit models, and that the use of
% bias-reducing adjustments to the score equations is an appealing
% alternative strategy.

\subsection{Omitting regressors affected by separation}

\cite{bias:SantosSilva+Tenreyro:2010} show that the regressors
responsible for separation in Poisson models can be easily identified
by running a least squares regression on the non-boundary observations
and checking for perfect collinearity among the regressors. The same
strategy is also applicable for Tobit regression models.

Having identified the collinear regressors associated with
separation, \cite{bias:SantosSilva+Tenreyro:2010} propose to simply
omit those and re-estimate the model using the \emph{full} data set with
all $n$~observations. The same strategy is also adopted in
\citet[][Chapter~6.2]{bias:Cameron+Trivedi:2013}, who suggest to drop
the separating regressor from the binary model part of a count data
hurdle model.

However, this strategy only leads to consistent estimates if the
omitted regressors are, in fact, not relevant, or were constructed to
specifically indicate a zero response \citep[e.g., in the artificial
data set used in the illustrations
of][]{bias:SantosSilva+Tenreyro:2011}. In contrast, when a highly
informative regressor is omitted, separation will be replaced by a
systematic misspecification of the model
\citep{bias:Heinze+Schemper:2002, bias:Zorn:2005}.  In that situation,
consistent estimates can be obtained by not only omitting the
regressor but also the observations responsible for separation, i.e.,
considering only the first term (\ref{loglik1}) in the likelihood and
dropping~(\ref{loglik2}).

% The above discussion highlights the appeal of using an alternative to
% the ML estimator that takes finite values regardless if separation
% occurs or not, allowing to keep all the regressors in the model.

\subsection{Bias reduction}

\citet{bias:Kosmidis+Firth:2020} have formally shown that, in
logit
%stats% logistic
regression models with full-rank model matrix, the bias-reduced (BR)
estimators coming from the adjusted score equations in
\citet{bias:Firth:1993} (i)~have always finite value and (ii)~shrink
towards zero in the direction of maximizing the Fisher information
about the parameters. There are also strong empirical findings that
the finiteness of the BR estimator extends beyond
logit models.
%stats% logistic regression.

A desirable feature of the bias-reducing adjustments to the score
functions is that they are asymptotically dominated by the score
functions. As a result, inference that relies on the BR estimates
(Wald tests, information criteria, etc.) can
be performed as usual by simply using the BR estimates in place of the
ML estimates. This makes BR estimation a rather attractive alternative
approach for dealing with separation, without omitting regressors.

While bias reduction is a well-established remedy for data separation
in
binary
%stats% binomial
regression models, it is less well known that it is
effective also in more general settings such as generalized nonlinear
models \citep{bias:Kosmidis+Firth:2009}, and, as illustrated here, the
models in Section~\ref{sec:intro}.

\section{Illustration}

Similarly to \cite{bias:SantosSilva+Tenreyro:2011}, we consider models
with intercept $x_{i, 1} = 1$ and regressors $x_{i, 2}$ and
$x_{i, 3}$ $(i = 1, \ldots, n)$. The values for $x_{i, 2}$ are
generated from a uniform distribution as
$x_{i,2} \sim \mathcal{U}(-1, 1)$. The values for $x_{i, 3}$
are, then, generated from Bernoulli distributions as
$x_{i,3} \sim \mathcal{B}(\pi)$ if $x_{i,2} > 0$ and
$x_{i,3} \sim \mathcal{B}(1 - \pi)$ otherwise, in order to allow for
correlation between the two regressors.

The responses for the Poisson model are generated
from (\ref{dgp1}) using $h(x_i^\top \beta) = \exp(x_i^\top \beta)$ and
the Poisson distribution for $F$ (with known dispersion $\phi =
1$). The responses for the Tobit model are generated
from a latent normal distribution $\mathcal{N}(x_i^\top\beta, \phi)$
with variance $\phi = 2$ and subsequent censoring by setting
all negative responses to~$0$.

For illustration purposes, we generate a single artificial data set
involving $n = 100$ regressor values with $\pi = 0.25$, and Poisson
and Tobit responses using $\beta_1 = 1$, $\beta_2 = 1$ and
$\beta_3 = -10$. In both cases, separation occurs due to the extreme
value for the coefficient of $x_{i, 3}$.  In the Appendix, we carry out
a thorough simulation study with $10{,}000$ data sets for a range of
combinations of $n$ and $\pi$ and $\beta_2 = -3$ so that separation
occurs with smaller probability.

We estimate the models from the artificial data using ML and BR
estimation using all $n = 100$ observations, and ML estimation of the
reduced model after omitting $x_{i,3}$ either by using just the subset
of the data set with $x_{i,2} = 0$ (ML/sub), or all $n = 100$
observations as proposed by \cite{bias:SantosSilva+Tenreyro:2010}
(ML/SST).

The bias-reducing adjusted score equations for the Poisson regression
are $\sum_{i = 1}^n (y _i + h_i /2 - \mu_i) x_i = 0_p$, where $0_p$ is
a $p$-vector of zeros and $h_i = x_i^\top (X^\top W X)^{-1} x_i \mu_i$
with $W = \mathop{\rm diag} \{\mu_1, \ldots, \mu_n\}$
\citep{bias:Firth:1992}. It is solved with the \texttt{brglm\_fit} method
from the \textsf{R} package \emph{brglm2} \citep{bias:brglm2}. For the
Tobit model we derived the adjusted score equations along with an
implementation in the \textsf{R} package  \emph{brtobit} \citep{bias:brtobit}.
The derivations are tedious but not complicated and are provided in the Appendix.

\begin{table}[t!]
\centering
\small
\caption{Comparison of different approaches when
  dealing with separation in a Poisson model.
  N is the number of observations used.\label{tab:poisson}}
\begin{tabular}{lD{.}{.}{3}D{.}{.}{3}D{.}{.}{3}D{.}{.}{3}}
\toprule
 & 
\multicolumn{1}{c}{ML} & 
\multicolumn{1}{c}{BR} & 
\multicolumn{1}{c}{ML/sub} & 
\multicolumn{1}{c}{ML/SST}\\
\midrule
(Intercept)    &     0.951  &  0.958  &  0.951  &  0.350 \\
               &    (0.100) & (0.099) & (0.100) & (0.096)\\
x2             &     1.011  &  1.006  &  1.011  &  1.662 \\
               &    (0.158) & (0.157) & (0.158) & (0.144)\\
x3             &   -20.907  & -5.174  &         &        \\
               & (2242.463) & (1.416) &         &        \\
\midrule
Log-likelihood & -107.364 & -107.869 & -107.364 & -169.028\\
N              &  100     &  100     &   55     &  100    \\
\bottomrule
\end{tabular}\end{table}

\begin{table}[t!]
\centering
\small
\caption{Comparison of different approaches when
  dealing with separation in a Tobit model.
  N is the number of observations used.\label{tab:tobit}}
\begin{tabular}{lD{.}{.}{3}D{.}{.}{3}D{.}{.}{3}D{.}{.}{3}}
\toprule
 & 
\multicolumn{1}{c}{ML} & 
\multicolumn{1}{c}{BR} & 
\multicolumn{1}{c}{ML/sub} & 
\multicolumn{1}{c}{ML/SST}\\
\midrule
(Intercept)    &      1.135  &  1.142  &  1.135  & -0.125 \\
               &     (0.208) & (0.210) & (0.208) & (0.251)\\
x2             &      0.719  &  0.705  &  0.719  &  2.074 \\
               &     (0.364) & (0.359) & (0.364) & (0.404)\\
x3             &    -11.238  & -4.218  &         &        \\
               & (60452.270) & (0.891) &         &        \\
(Variance)     &      1.912  &  1.970  &  1.912  &  3.440 \\
               &     (0.422) & (0.434) & (0.422) & (0.795)\\
\midrule
Log-likelihood & -87.633 & -88.101 & -87.633 & -118.935\\
N              & 100     & 100     &  55     &  100    \\
\bottomrule
\end{tabular}\end{table}  

Tables~\ref{tab:poisson} and~\ref{tab:tobit} show the results from
estimating the Poisson and Tobit models, respectively, with the four
different strategies. The following remarks can be made:
\begin{itemize}

\item Standard ML estimation using all observations leads to a large
  estimate of $\beta_3$ with even larger standard error. As a result,
  a standard Wald test results in no evidence against the hypothesis
  that $x_3$ should not be in the model, despite the fact that using
  $\beta_3 = -10$ when generating the data makes $x_3$ perhaps the
  most influential regressor.\footnote{The estimates for $\beta_3$ and
    the corresponding standard errors are formally
    infinite. The displayed finite values are the result of stopping
    the iterations early according to the convergence criteria used
    during maximization of the likelihood. Stricter convergence
    criteria will result in estimates and standard errors
    that diverge further.} 

\item The ML/sub strategy, i.e., estimating the model without $x_2$
  only for the 0~observations with $x_{i,2} =
  0$, yields exactly the same estimates as ML because it optimizes the
  term (\ref{loglik1}), after setting (\ref{loglik2}) to zero.

\item Compared to ML and ML/sub, BR has the advantage of returning a
  finite estimate and standard error for
  $\beta_3$. Hence a Wald test can be directly used to examine the
  evidence against
  $\beta_3~=~0$.  The other parameter estimates and the log-likelihood
  are close to ML. Similarly to binary response models, bias reduction
  here slightly shrinks the parameter estimates of
  $\beta_2$ and $\beta_3$ towards zero.

\item Finally, the estimates from ML/SST, where regressor $x_3$ is
  omitted and all observations are used, appear to be far from the values
  we used to generate the data. This is due to the fact that $x_3$ is
  not only highly informative but also correlated
  with~$x_2$.
\end{itemize}
Moreover, the simulation experiments in the Appendix provide evidence that the
BR estimates are always finite, and result in Wald-type intervals with
better coverage.

\bibliographystyle{elsarticle-harv}
\bibliography{brtobit}

\clearpage

\begin{appendix}
  \begin{onecolumn}

\section{Bias-reducing adjusted score functions for Tobit regression}

The Tobit model is one of the classic models of microeconometrics. Fundamental results were obtained by \cite{bias:Amemiya:1973}. A detailed account of basic properties is available in, e.g., \cite{bias:Gourieroux:2000}. Here we provide the building blocks for bias-reduced estimation of the Tobit model. 

Denote by $\ell(\theta)$ the log-likelihood function for a Tobit
regression model with full-rank, $n \times p$ model matrix $X$ with
rows the $p$-vectors $x_1, \ldots, x_n$, and a $(p+1)$-vector of
parameters $\theta = (\beta^\top, \phi)^\top$ with regression
parameters $\beta$ and variance $\phi$. Then,
$\ell(\theta) = \sum_{i = 1}^n [ (1 - d_i) \log (1 - F_i) - d_i (\log
\phi) / 2 - d_i (y_i - \eta_i)^2 / (2 \phi) ]$, where $d_i = 1$ if
$y_i > 0$ and $d_i = 0$ if $y_i \leq 0$, $\eta_i = x_i^\top\beta$, and $F_i$ is the standard normal
distribution function at $\eta_i/\sqrt{\phi}$. The score vector is
\[
s(\theta) = \nabla \ell(\theta) = 
  \begin{bmatrix}
   s_\beta(\theta) \\
   s_\phi(\theta)
  \end{bmatrix}
  =
  \begin{bmatrix}
    \displaystyle \sum_{i = 1}^n \left\{ \frac{(d_i - 1) \lambda_i}{\sqrt{\phi}}  + \frac{d_i ( y_i - \eta_i )}{\phi}  \right\} x_i \\   
    \displaystyle \sum_{i = 1}^n \left\{ \frac{(1 - d_i) \lambda_i \eta_i}{2 \phi^{3/2}} - \frac{d_i}{2 \phi} + \frac{d_i (y_i - \eta_i)^2}{2\phi^2}\right\}
  \end{bmatrix}\, ,  
\]
where $\lambda_i = f_i / (1 - F_i)$ and $f_i$ is the density function of the standard normal distribution at $\eta_i/\sqrt{\phi}$. 

The observed information matrix, $j(\theta) = - \nabla\nabla^\top \ell(\theta)$, has the form
\[
  j(\theta) =
  \begin{bmatrix}
   j_{\beta\beta}(\theta) & j_{\beta\phi}(\theta) \\
   j_{\phi\beta}(\theta) & j_{\phi\phi}(\theta)
  \end{bmatrix} \, ,
\]
where, setting $\nu_i = f_i / (1 - F_i)^2$, 
\begin{eqnarray*}
  j_{\beta\beta}(\theta) & = & \sum_{i = 1}^n\left[ \frac{\nu_i(d_i - 1)}{\sqrt{\phi}} \left\{ \frac{f_i}{\sqrt{\phi}} - \frac{(1- F_i) \eta_i}{\phi}\right\} - \frac{d_i}{\phi} \right] x_i x_i^\top \, ,\\ 
  j_{\beta\phi}(\theta) & = & \sum_{i = 1}^n\left[
                      \frac{\nu_i (d_i - 1)}{2 \phi^{3/2}} \left\{
                      \frac{(1  -F_i)\eta_i^2}{\phi} - 1 + F_i - \frac{\eta_if_i}{\sqrt{\phi}}
                      \right\} - \frac{d_i (y_i - \eta_i)}{\phi^2}
                              \right] x_i \,,\\  %^\top
  j_{\phi\beta}(\theta) & = & j_{\beta\phi}(\theta)^\top \,,\\
j_{\phi\phi}(\theta) & = &  \sum_{i = 1}^n \left[\frac{\nu_i(1 - d_i) }{4 \phi^{5/2}} \left\{ \frac{(1 - F_i) \eta_i^3}{\phi} - 3 ( 1- F_i) \eta_i - \frac{\eta_i^2 f_i}{\sqrt{\phi}}\right\} + \frac{d_i}{2\phi^2} - \frac{d_i (y_i - \eta_i)^2}{\phi^3} \right] \, .
\end{eqnarray*}

As shown in \citet{bias:Kosmidis+Firth:2010}, a BR estimator for
$\theta$ results as the solution of the adjusted score equations
$s(\theta) + A(\theta) = 0_{p + 1}$, where the vector $A(\theta)$ has $t$-th
component
$A_t(\theta) = \tr[\{i(\theta)\}^{-1} \{P_t(\theta) + Q_t(\theta)\}] /
2$ $(t = 1, \ldots, p + 1)$.  In the above expression,
$Q_t(\theta) = - \expect(j(\theta)s_t(\theta))$ and
$P_t(\theta) = \expect(s(\theta)s^\top(\theta)s_t(\theta))$, where
$i(\theta) = \expect(j(\theta))$ is the expected
information matrix. The \textsf{R} package
\emph{brtobit} implements $i(\theta)$, $Q_t(\theta)$, and
$P_t(\theta)$, and solves the bias-reducing adjusted score equations
for general Tobit regressions using the quasi Fisher-scoring scheme
proposed in \citet{bias:Kosmidis+Firth:2010}.

The matrices $i(\theta)$, $Q_t(\theta)$ and $P_t(\theta)$ have the
same block structure as $j(\theta)$ and, directly by their
definition, closed-form expressions for their blocks result by taking
expectations of the appropriate products of blocks of $s(\theta)$ and
$j(\theta)$. By direct inspection of the expressions for $s(\theta)$
and $j(\theta)$, the required expectations result by noting that
$\expect(d_i^m) = F_i$, $\expect((1 - d_i)^m) = 1 - F_i$,
$\expect(d_i^m(1 - d_i)^l) = 0$, $\expect(d_i^m(1 - d_i)^l (y_i - \eta_i)^k) = 0$, and by computing
$\expect(d_i^m (y_i - \eta_i)^l)$ $(k, l, m = 1, \ldots, 6)$. For the
latter expression, note that
$\expect(d_i^m (y_i - \eta_i)^l) = F_i \expect((y_i - \eta_i)^l \mid
y_i > 0)$, and that some algebra gives
\begin{eqnarray*}
  \expect(y_i - \eta_i \mid y_i > 0) & = & \sqrt{\phi} \xi_i \, , \\
  \expect((y_i - \eta_i)^2 \mid y_i > 0) & = & \phi - \sqrt{\phi} \eta_i \xi_i \,, \\
  \expect((y_i - \eta_i)^3 \mid y_i > 0) & = & \sqrt{\phi} \xi_i (\eta_i^2 + 2 \phi) \,, \\
  \expect((y_i - \eta_i)^4 \mid y_i > 0) & = & 3\phi^2 - \eta_i^3\sqrt{\phi} \xi_i - 
                                                             3 \phi^{3/2}\eta_i\xi_i \,, \\
  \expect((y_i - \eta_i)^5 \mid y_i > 0) & = & \sqrt{\phi}\eta_i^4\xi_i + 4\phi^{3/2}\xi_i(\eta_i^2 + 2\phi) \, , \\
  \expect((y_i - \eta_i)^6 \mid y_i > 0) & = & -\eta_i\sqrt{\phi}\xi_i(\eta_i^4 + 
                                                             5\eta_i^2\phi + 15\phi^2) + 15\phi^3 \, ,
\end{eqnarray*}
where $\xi_i = f_i/F_i$. 
The expected information, 
\[
 i(\theta) =
  \begin{bmatrix}
   \expect(j_{\beta\beta}(\theta)) & \expect(j_{\beta\phi}(\theta)) \\
   \expect(j_{\phi\beta}(\theta)) & \expect(j_{\phi\phi}(\theta))
  \end{bmatrix},
\]  
has elements 
\begin{eqnarray*}
\expect(j_{\beta\beta}(\theta)) & = & -\frac{1}{\phi} \sum_{i = 1}^n \left\{\frac{\eta_if_i}{\sqrt{\phi}} - \lambda_if_i - F_i\right\}x_i x_i^\top \, , \\
\expect(j_{\beta\phi}(\theta)) & = & \frac{1}{2\phi^{3/2}}\sum_{i = 1}^nf_i\left\{\frac{\eta_i^2}{\phi} + 1 - \lambda_i\frac{\eta_i}{\sqrt{\phi}}\right\}x_i^\top \, , \\
 \expect(j_{\phi\beta}(\theta)) & = & \expect(j_{\beta\phi}(\theta))^\top \, , \\
\expect(j_{\phi\phi}(\theta)) & = & -\frac{1}{4\phi^2}\sum_{i = 1}^n\left\{f_i\frac{\eta_i^3}{\phi^{3/2}} + f_i\frac{\eta_i}{\sqrt{\phi}} - \lambda_if_i\frac{\eta_i^2}{\phi} - 2 F_i \right\}. 
\end{eqnarray*}         
Furthermore, for $t \in \{1, \ldots, p\}$,
\[
Q_t(\theta) =
- \begin{bmatrix}
 \expect(j_{\beta\beta} s_{\beta_t}) & \expect(j_{\beta\phi} s_{\beta_t}) \\
 \expect(j_{\beta\phi} s_{\beta_t})^\top & \expect(j_{\phi\phi} s_{\beta_t})
\end{bmatrix}
\quad \text{and} \quad
P_t(\theta) =
 \begin{bmatrix}
 \expect(s_{\beta}s_{\beta}^\top s_{\beta_t}) & \expect(s_{\beta}s_{\phi} s_{\beta_t}) \\
 \expect(s_{\beta}s_{\phi} s_{\beta_t})^\top & \expect(s_{\phi}s_{\phi}s_{\beta_t})
 \end{bmatrix}\, ,
\]
and for $t = p + 1$, 
\[
Q_{p + 1}(\theta) = 
- \begin{bmatrix}
 \expect(j_{\beta\beta} s_\phi) & \expect(j_{\beta\phi} s_\phi) \\
 \expect(j_{\beta\phi} s_\phi)^\top & \expect(j_{\phi\phi} s_\phi)
\end{bmatrix}
\quad \text{and} \quad
P_{p + 1}(\theta) = 
 \begin{bmatrix}
 \expect(s_{\beta}s_{\beta}^\top s_{\phi}) & \expect(s_{\beta}s_{\phi}s_{\phi}) \\
 \expect(s_{\beta}s_{\phi}s_{\phi})^\top & \expect(s_{\phi}s_{\phi}s_{\phi})
 \end{bmatrix},
\]
where
\begin{eqnarray*}
 \expect(j_{\beta\beta} s_{\beta_t}) & = & \sum_{i = 1}^n\left[-\frac{f_i}{\phi^{3/2}}\left(\lambda_i^2 - \frac{\lambda_i \eta_i}{\sqrt{\phi}} -1 \right)\right]x_i x_i^\top x_{i,t}\,, \\
 \expect(j_{\beta\phi} s_{\beta_t}) & = & \sum_{i = 1}^n\left[ \frac{1}{2\phi^2}\lambda_i f_i \left\{-\frac{\eta_i^2}{\phi} + 1 + \lambda_i\frac{\eta_i}{\sqrt{\phi}} \right\} + \frac{1}{\phi^2}\left\{F_i - \frac{\eta_i f_i}{\sqrt{\phi}} \right\} \right]x_i^\top x_{i,t}\,, \\
  \expect(j_{\phi\phi} s_{\beta_t}) & = & \sum_{i = 1}^n \frac{1}{\phi^{5/2}}\left[\lambda_i\frac{f_i \eta_i}{4 \sqrt{\phi}} 
 \left\{\frac{\eta_i^2}{\phi} - 3 - \lambda_i \frac{\eta_i}{\sqrt{\phi}} \right\} + \frac{f_i \eta_i^2}{\phi} + \frac{3 f_i}{2} \right]x_{i,t}\,, \\
 \expect(j_{\beta\beta} s_\phi) & = & \sum_{i = 1}^n\left[ \frac{f_i^2 \eta_i}{2\phi^{5/2}(1 - F_i)} \left\{\lambda_i - \frac{\eta_i}{\sqrt{\phi}}  \right\} - \frac{\eta_i f_i}{2 \phi^{5/2}} \right]x_i x_i^\top\,, \\
 \expect(j_{\beta\phi} s_\phi) & = & \sum_{i = 1}^n\left[ \lambda_i \frac{f_i \eta_i}{4\phi^3} \left\{\frac{\eta_i^2}{\phi} - 
            1 - \lambda_i\frac{\eta_i}{\sqrt{\phi}} \right\} + \frac{f_i}{2\phi^{5/2}} \left\{1 + \frac{\eta_i^2}{\phi}\right\} \right]x_i^\top\,, \\ 
 \expect(j_{\phi\phi} s_\phi) & = & \sum_{i = 1}^n\left[ \lambda_i\frac{\eta_i^2}{8\phi^4} \left\{-\frac{\eta_i^2 f_i}{\phi} + 3f_i + \lambda_i \frac{f_i \eta_i}{\sqrt{\phi}}\right\} + \frac{F_i}{\phi^3} - \frac{3\eta_i f_i}{4 \phi^{7/2}} - \frac{f_i \eta_i^3}{2\phi^{9/2}} \right] \,,\\
  \expect(s_{\beta}s_{\beta}^\top s_{\beta_t}) & = & \sum_{i = 1}^n\left[ -\lambda_i^2\frac{f_i}{\phi^{3/2}} + \frac{f_i}{\phi^{5/2}} \left\{ 
            \eta_i^2 + 2\phi\right\} \right]x_i x_i^\top x_{i,t} \,, \\
 \expect(s_{\beta}s_{\beta}^\top s_{\phi}) & = & \sum_{i = 1}^n\left[ \frac{\eta_i f_i}{2\phi^{5/2}} \left\{\lambda_i^2 - 2 - \frac{\eta_i^2}{\phi}\right\} +\frac{F_i}{\phi^2} \right]x_i x_i^\top \,,\\
 \expect(s_{\beta}s_{\phi}s_{\beta_t}) & = & \sum_{i = 1}^n\left[\frac{f_i \eta_i}{2 \phi^{5/2}} \left\{ \lambda_i^2 - 2 \right\} + \frac{F_i}{\phi^2} - \frac{f_i \eta_i^3}{2\phi^{7/2}} \right]x_i x_{i,t} \,,\\
 \expect(s_{\beta}s_{\phi}s_{\phi}) & = & \sum_{i = 1}^n\left[\frac{f_i \eta_i^2}{2\phi^{7/2}} \left\{-\lambda_i^2\frac{1}{2} + 1 \right\} +
                                          \frac{f_i}{4\phi^{5/2}} \left\{5 + \frac{\eta_i^4}{\phi^2}\right\} \right]x_i \,,\\
  \expect(s_{\phi}s_{\phi}s_{\beta_t}) & = & \sum_{i = 1}^n\left[ -\frac{f_i \eta_i^2}{4\phi^{7/2}} \left\{ \lambda_i^2 - 
            2 - \frac{\eta_i^2}{\phi}\right\} + \frac{5 f_i}{4 \phi^{5/2}} \right]x_{i,t} \,,\\
  \expect(s_{\phi}s_{\phi}s_{\phi}) & = & \sum_{i = 1}^n\left[ \frac{f_i\eta_i^3}{8\phi^{9/2}} \left\{\lambda_i^2 - 2 - \frac{\eta_i^2}{\phi} \right\} + \frac{F_i}{\phi^3} - \frac{9f_i\eta_i}{8\phi^{7/2}} \right].
\end{eqnarray*}

\section{Simulation}

The aim of the simulation experiment is to compare the performance of
the BR and ML estimator in count and limited dependent variable models
with varying probabilities of infinite ML estimates. The comparison
here is in terms of bias, variance, and empirical coverage of nominally
$95\%$ Wald-type confidence intervals based on the asymptotic
normality of the estimators. Our results were obtained using 
\textsf{R}~4.0.3 \citep{bias:R:2020}. 
Random variables were generated using the default methods for the relevant distributions, which in turn rely on uniform random numbers obtained by the Mersenne
Twister, currently \textsf{R}'s default generator.

The same data generating process as in Section 3 of the main paper is
considered, with the coefficient of the binary regressor $x_2$ set to
the less extreme value $\beta_3 = -3$. The amount of correlation
between $x_2$ and $x_3$ varies with
$\pi \in \{0, 1/8, 1/4, 3/8, 1/2\}$ so that increasing the value of
$\pi$ leads to decreasing the probability of infinite estimates. The
sample sizes we consider are $n \in \{25, 50, 100, 200, 400\}$. For
each combination of $\pi$ and $n$, $10{,}000$ independent samples are
simulated, and the parameters of the Poisson and Tobit regression
models in Section 3 are estimated using maximum likelihood and bias
reduction. The estimates are then used to compute simulation-based
estimates of the bias, variance, and coverage probability for
$\beta_3$.

For the ML estimator, the bias, variance, and coverage probabilities
are computed conditionally on the finiteness of the ML estimates. We
classify an ML estimate as infinite if the corresponding estimated
standard error exceeds $20$. In effect, we are assuming that if the
standard error exceeds $20$, the Fisher scoring iteration for ML
stopped while moving along an asymptote on the log-likelihood surface,
hence, at a point where the inverse negative hessian has at least one
massive diagonal element. The heuristic value $20$ is conservative
even for $n = 25$. This has been verified through a pilot simulation
study to estimate the variance of the reduced-bias estimator, which
has the same asymptotic distribution as the ML estimator. No
convergence issues were encountered and the maximum estimated
standard error of the reduced-bias estimators accross simulation
settings, parameters, and sample sizes was $8.3$ for Tobit and $5.5$
for Poisson regression.

For BR estimation, the estimates appear to be always finite. So, we
estimate biases, variances and coverage probabilities both
conditionally on the finiteness of the ML estimates and
unconditionally. We note here that a direct comparison of conditional
and unconditional summaries is not formally valid, but gets more and
more informative as the probability of infinite estimates decreases.

Figures~\ref{fig:sim-05}, \ref{fig:sim-06}, \ref{fig:sim-07}, and
\ref{fig:sim-08} show the estimated probability that the ML and BR
estimate of $\beta_3$ are infinite, the estimated bias, the estimated
variance, and the estimated coverage probability of $95\%$ Wald-type
confidence intervals, respectively, for the Poisson
model. Figures~\ref{fig:sim-01}, \ref{fig:sim-02}, \ref{fig:sim-03},
and \ref{fig:sim-04} show the corresponding results for the Tobit
model.

The results for Poisson and Tobit regression lead to similar insights:
\begin{itemize}

\item Bias reduction via adjusted score functions always yields finite
  estimates.

\item The BR estimator has bias close to zero even for small sample
  sizes.

\item Wald-type confidence intervals based on BR estimates have
  good coverage properties.

\item The variances of the BR and ML estimator get closer to each
  other and closer to zero as $n$ increases. This is exactly what the
  theory suggests because the score functions asymptotically
  dominate the bias-reducing adjustments.

\end{itemize}

\begin{figure}[p!]
\centering
\includegraphics[width=0.8\textwidth]{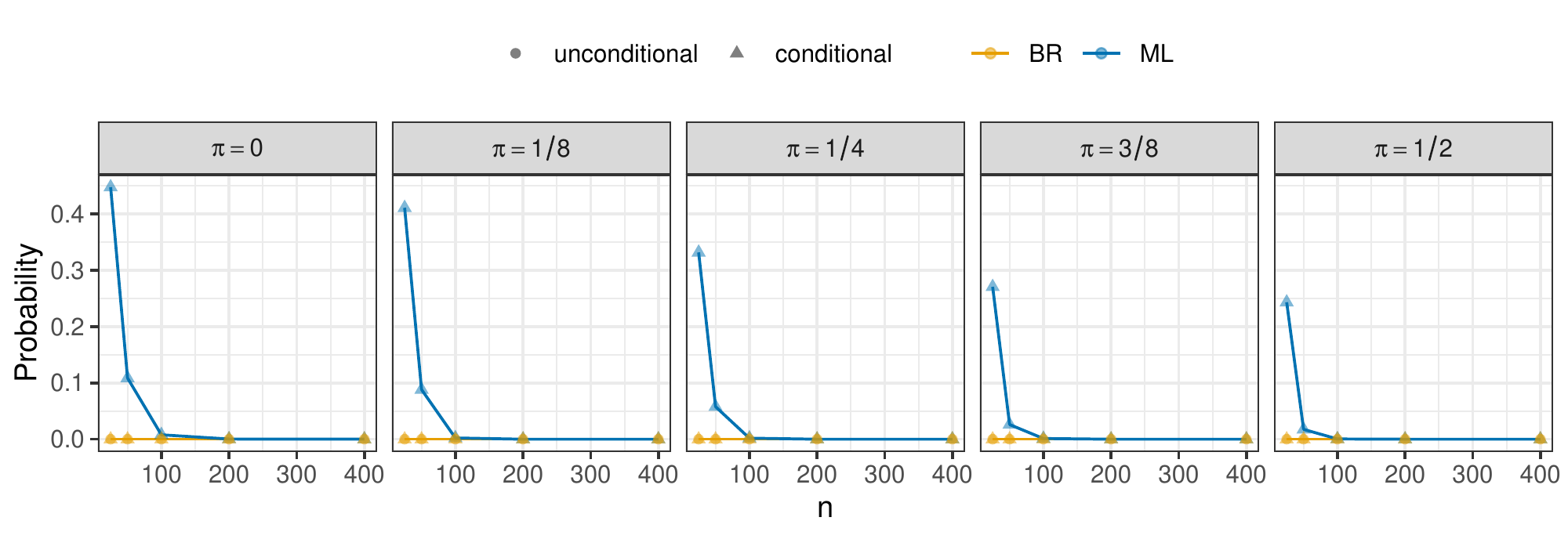}
\caption{\label{fig:sim-05} Probability of infinite estimates for $\beta_3$ (Poisson).}

\includegraphics[width=0.8\textwidth]{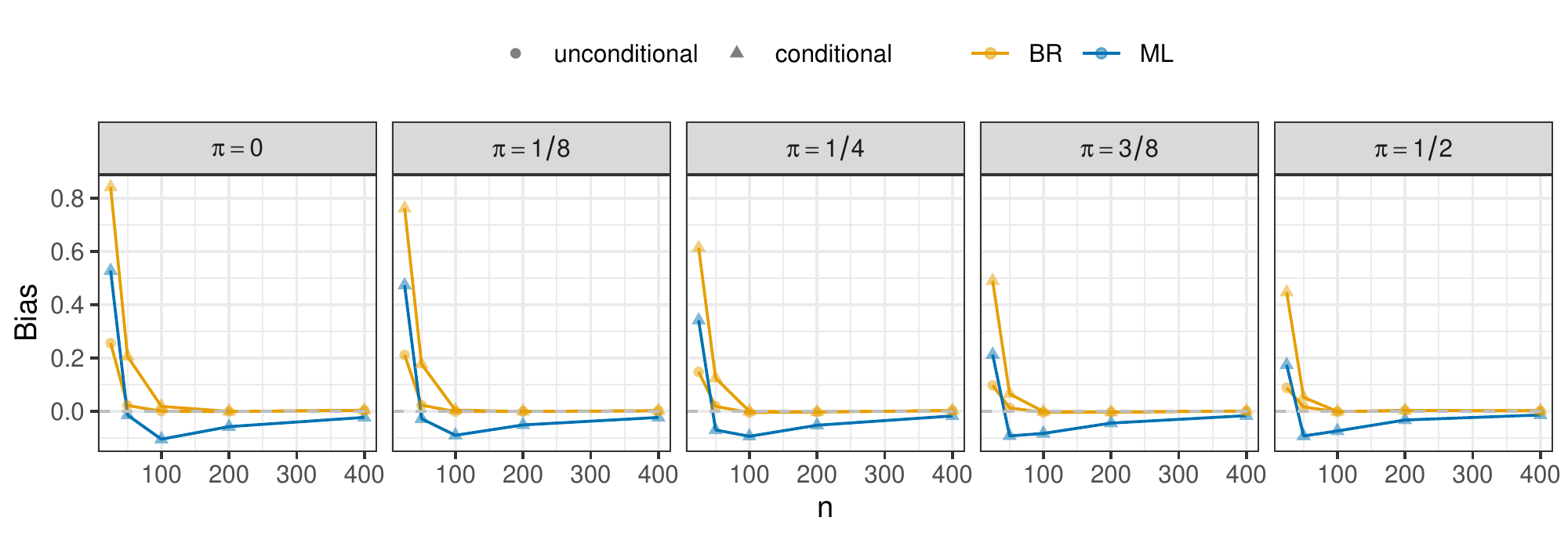}
\caption{\label{fig:sim-06} Bias of estimates for $\beta_3$ (Poisson).}

\includegraphics[width=0.8\textwidth]{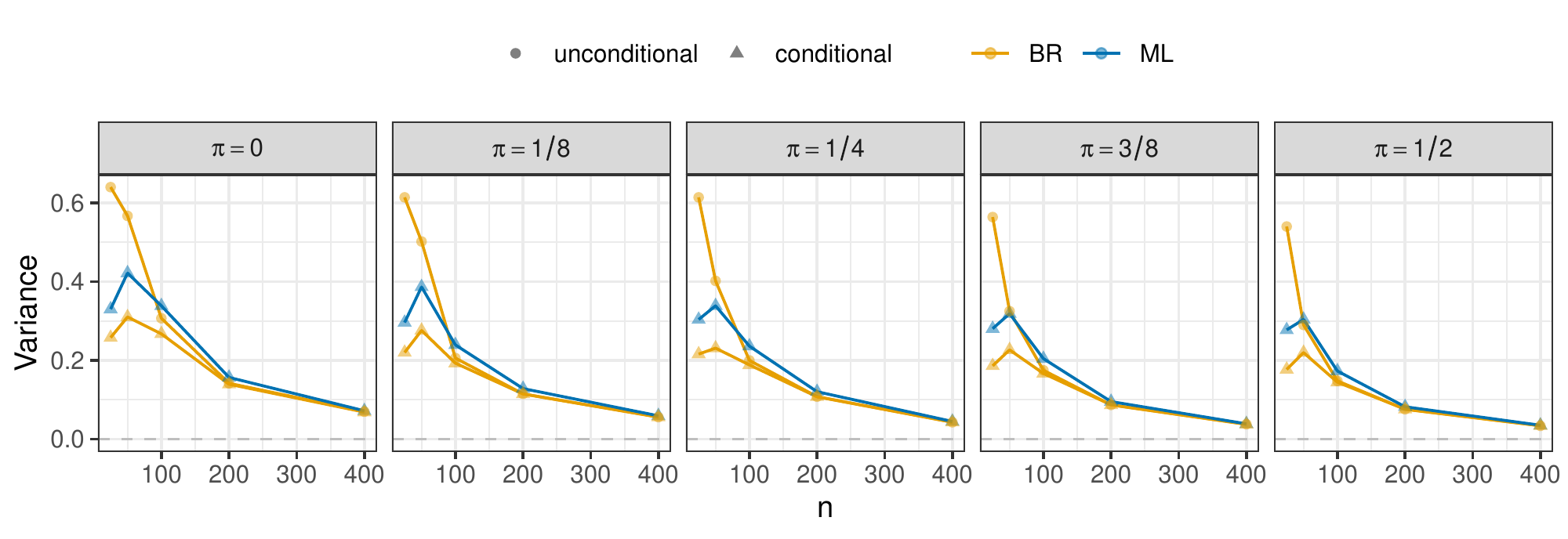}
\caption{\label{fig:sim-07} Variance of estimates for $\beta_3$ (Poisson).}

\includegraphics[width=0.8\textwidth]{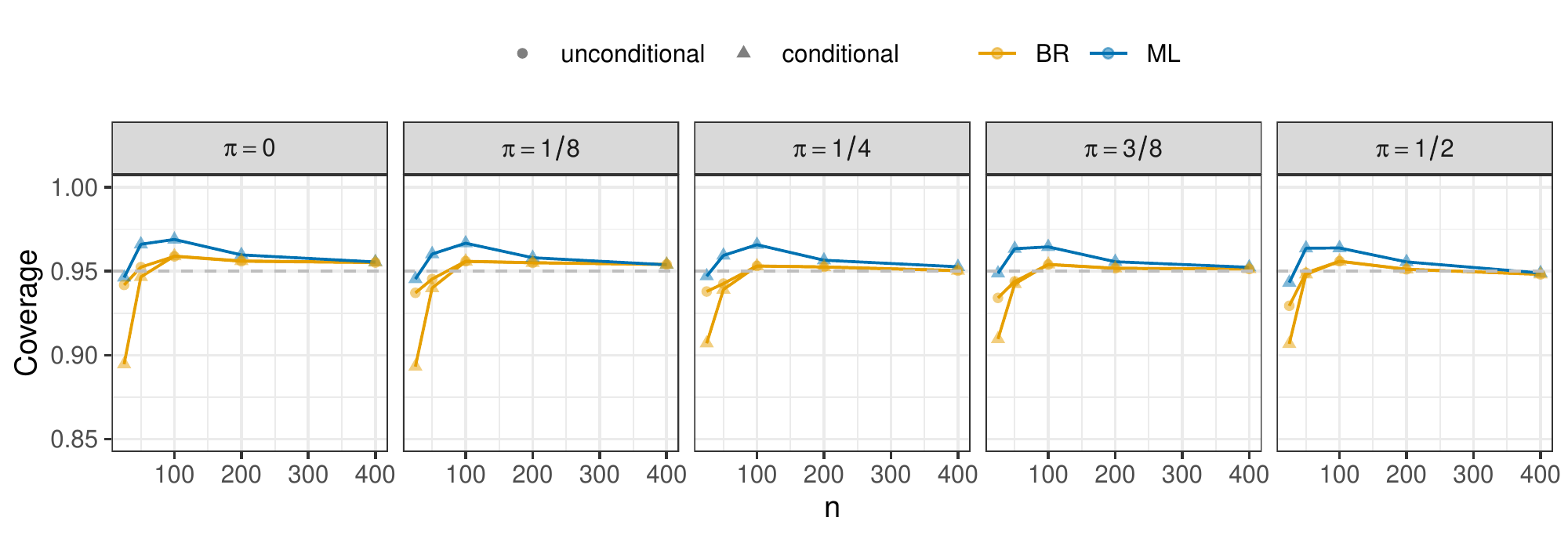}
\caption{\label{fig:sim-08} Coverage of $95\%$ Wald-type confidence intervals for $\beta_3$ (Poisson).}
\end{figure}

\begin{figure}[p!]
\centering
\includegraphics[width=0.8\textwidth]{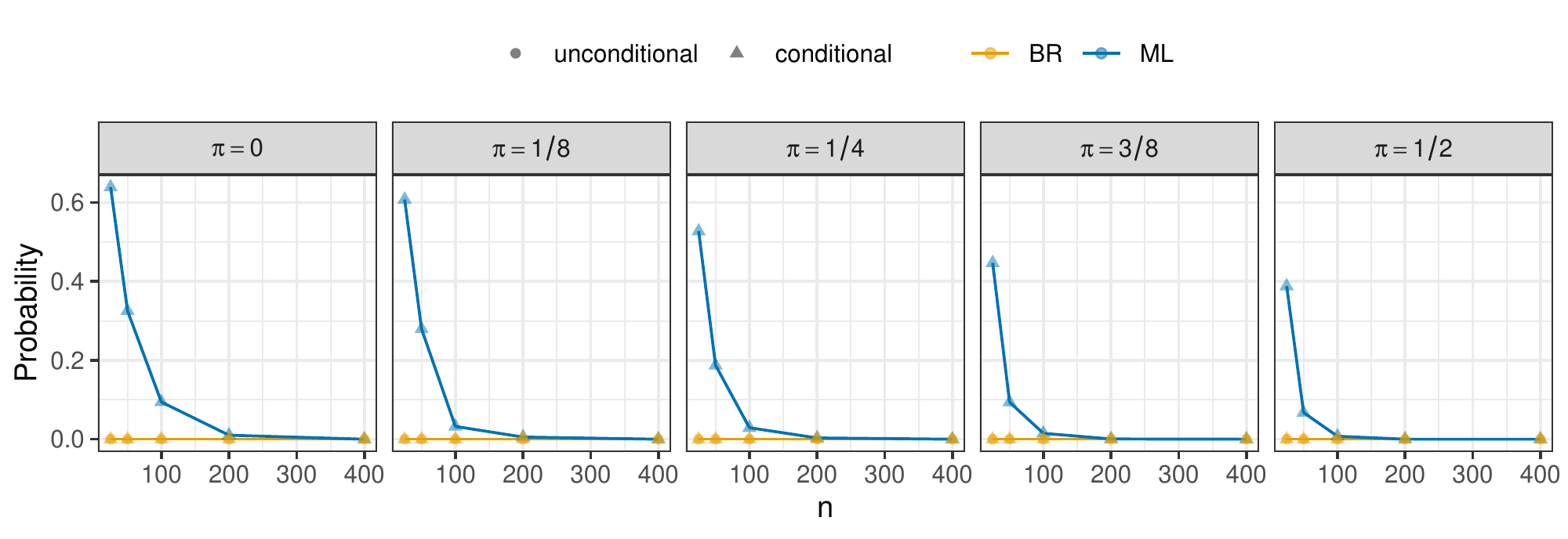}
\caption{\label{fig:sim-01} Probability of infinite estimates for $\beta_3$ (Tobit).}

\includegraphics[width=0.8\textwidth]{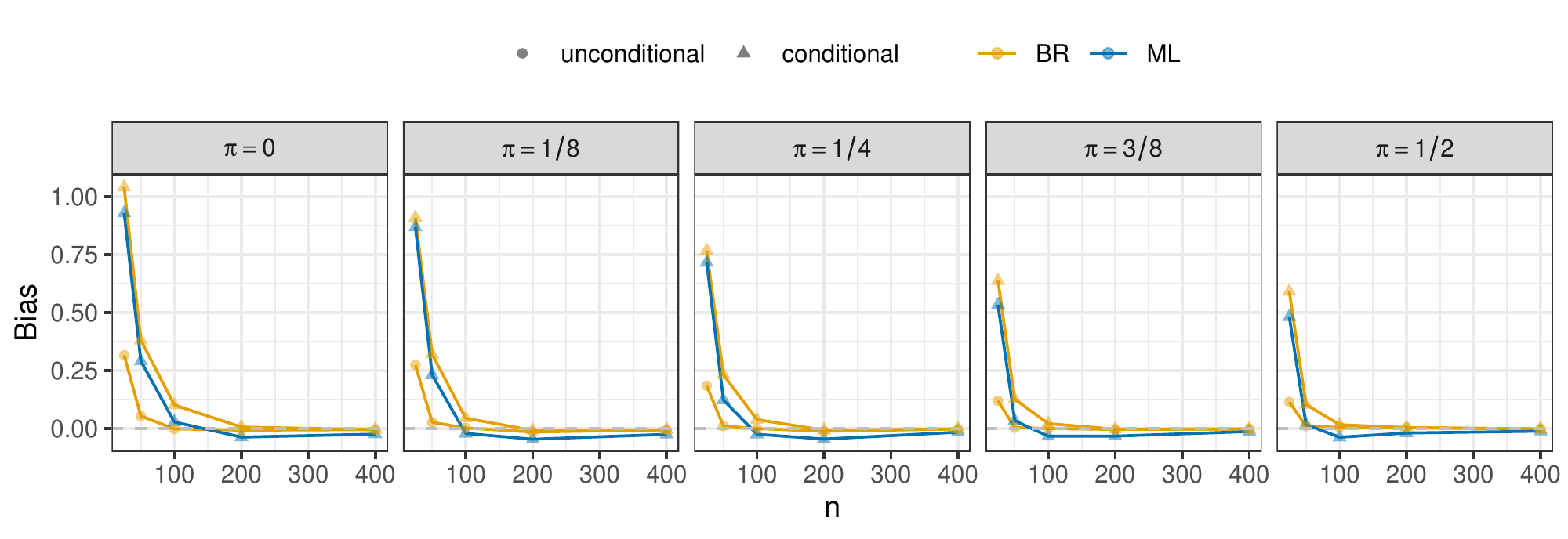}
\caption{\label{fig:sim-02} Bias of estimates for $\beta_3$ (Tobit).}

\includegraphics[width=0.8\textwidth]{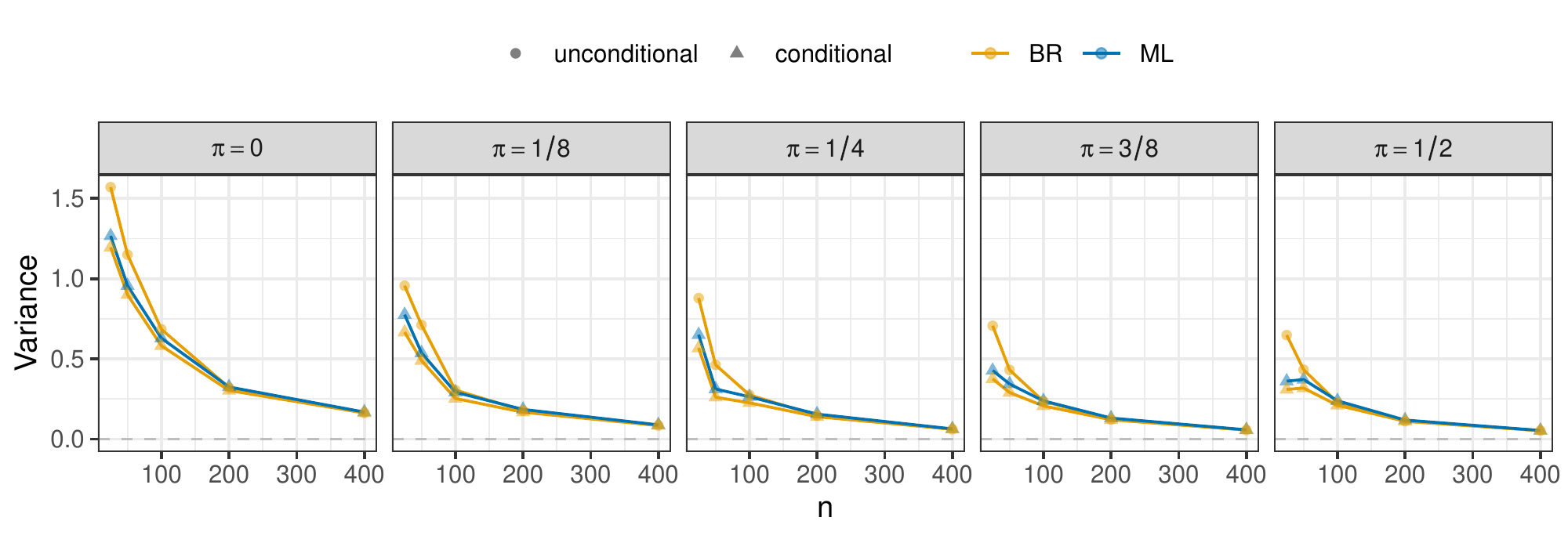}
\caption{\label{fig:sim-03} Variance of estimates for $\beta_3$ (Tobit).}

\includegraphics[width=0.8\textwidth]{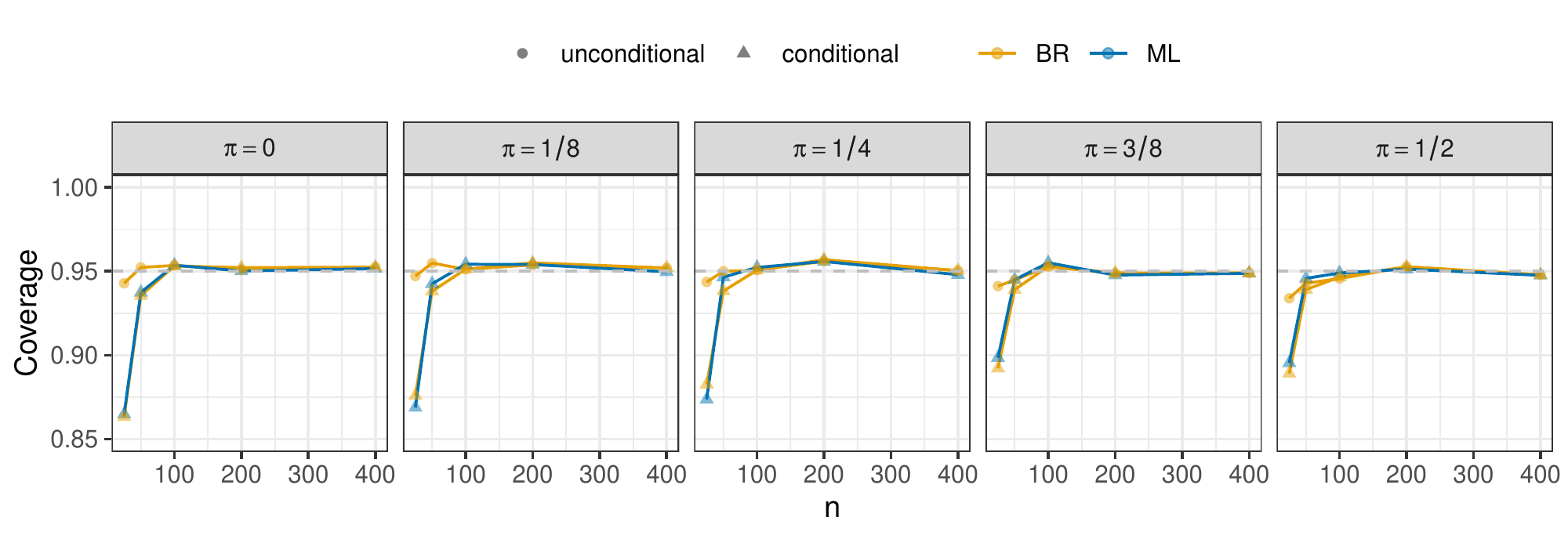}
\caption{\label{fig:sim-04} Coverage of $95\%$ Wald-type confidence intervals for $\beta_3$ (Tobit).}
\end{figure}

\end{onecolumn}
\end{appendix}

\end{document}